\documentclass[aps,prx,reprint,superscriptaddress,onecolumn,notitlepage]{revtex4-1}

\usepackage[a4paper,left=2cm,right=2cm,top=3cm,bottom=3cm]{geometry}
\usepackage{soul}
\usepackage{relsize}
\usepackage{lmodern}
\usepackage{slantsc}
\usepackage{graphicx}
\usepackage{amsmath}
\usepackage{amssymb}
\usepackage{amsthm}
\usepackage{amsbsy}
\usepackage{mathrsfs}
\usepackage{varioref}
\usepackage{dsfont}
\usepackage{bm}
\usepackage{color}
\usepackage[usenames,dvipsnames]{xcolor}
\usepackage[colorlinks = false, citecolor=green, linkcolor=red, breaklinks=true, linktocpage=true]{hyperref}
\usepackage[font = footnotesize, labelfont = bf, justification = RaggedRight]{caption}
\usepackage{subfigure}
\usepackage{booktabs} % for much better looking tables
\usepackage{array} % for better arrays (eg matrices) in maths
\usepackage{paralist} % very flexible & customisable lists (eg. enumerate/itemize, etc.)
\usepackage{verbatim} % adds environment for commenting out blocks of text & for better verbatim
\edef\restoreparindent{\parindent=\the\parindent\relax}
\usepackage[parfill]{parskip}
\restoreparindent
\usepackage{natbib}

\newcommand{\sub}[1]{_{\!\mathsmaller{\, #1}}}
\newcommand{\eq}[1]{Eq.~\eqref{#1}}
\newcommand{\fig}[1]{Fig.~\ref{#1}}

\newcommand{\app}[1]{Appendix~(\ref{#1})}

\newcommand{\thmref}[1]{Theorem~\ref{#1}}

\newcommand{\<}{\langle}
\renewcommand{\>}{\rangle}
\newcommand{\ket}[1]{|{#1}\rangle}

\newcommand{\h}{{\mathcal{H}}}

\newcommand{\s}{{\mathcal{S}}}

\newcommand{\xx}{{\mathcal{X}}}

\renewcommand{\aa}{{\mathcal{A}}}
\newcommand{\bb}{{\mathcal{B}}}

\newcommand{\mm}{{\mathcal{M}}}

\newcommand{\ii}{{\mathcal{I}}}

\newcommand{\nn}{\mathcal{N}}

\newcommand{\one}{\mathds{1}}
\newcommand{\zero}{\mathds{O}}

\newcommand{\tr}{\mathrm{tr}}

\theoremstyle{plain}

\newtheorem{thm}{Theorem}
\begin{document}

\title{Symmetry constrained decoherence of conditional expectation values}

\author{M. Hamed Mohammady}
\affiliation{Department of Physics, Lancaster University, LA1 4YB, United Kingdom}

\author{Alessandro Romito}
\affiliation{Department of Physics, Lancaster University, LA1 4YB, United Kingdom}

%\date{\today}

\begin{abstract}
Conditional expectation values of quantum mechanical observables reflect unique non-classical correlations, and  are generally sensitive to decoherence. We consider the circumstances under which such sensitivity to decoherence is removed, namely, when the  measurement process is subjected to  conservation laws. Specifically, we address systems with additive conserved quantities and identify sufficient conditions for the system state such that its coherence plays no role in the conditional expectation values of observables that commute with the conserved quantity. We discuss our findings for a specific model where the system-detector coupling is given by the Jaynes-Cummings interaction, which is relevant to experiments tracking trajectories of qubits in cavities. Our results clarify, among others, the role of coherence in thermal measurements in current architectures for quantum thermodynamics experiments.

\end{abstract}

\maketitle

\section{Introduction}

Quantum measurements stochastically disturb the state of the measured system. The paradigmatic example is ideal projective measurements, wherein   the system ``collapses'' to the  eigenstates of the measured observable \cite{vonNeumann}. More generally, the measurement induced back action is characterised by completely positive, trace non-increasing maps, or quantum instruments, which can be given a Kraus decomposition \cite{Kraus1983}.  Such a state change will therefore allow us to define what the  expectation value of some other observable will be, conditional on observing a given measurement outcome. In order to define the conditional \emph{change} in the expectation value of an observable, however, we must have a method of defining what the conditional expectation value of an observable would be prior to observing a measurement outcome. As recently suggested in \cite{Mohammady2018a}, this quantity can be defined by the weak value \cite{Aharonov1988, Dressel2014}.   The weak value can be arbitrarily large, and even imaginary, which is incompatible with a probability distribution over the eigenvalues of the observable in question. 

Conditional expectation values of observables have been associated to physically significant quantities like tunnelling times \cite{Steinberg1995, Choi2013, Zilberberg2014, Romito2014}  or the spectroscopy of a wave function \cite{Lundeen2011}. The possibility of tracking systems along single quantum trajectories conditioned to measurements outcomes in optical \cite{Guerlin2007,Sayrin2011}  and more recently in solid state \cite{Vijay2012, Hatridge2013} systems, opens also the possibility to observe individual quantum trajectories \cite{Murch2013}, implement feedback control protocols  \cite{Vijay2012, Blok2014,DeLange2014}, determine weak values \cite{Groen2013,Campagne-Ibarcq2014}, produce deterministic entanglement \cite{Riste2013, Roch2014}, and realize Maxwell demons \cite{Naghiloo2018}. It is then possible to detect physical quantities along such quantum trajectories. Most prominently this idea has been used to define thermodynamic quantities along quantum trajectories based on the definition of conditional energy change along trajectories \cite{Alonso2016, Alexia-measurement-thermodynamics,Elouard2018}.

These conditional values are extremely sensitive to the coherence of the initial state of the system. This also means that they are extremely sensitive to decoherence mechanisms \cite{Romito2010a, Thomas2012}. In the case of the weak value, for example, the effect of decoherence has been studied by analysing the quantum operation acting on the W-operator \cite{weak-value-decoherence}.  However, the effect of decoherence can be constrained in some situations by the presence of symmetries, i.e. conserved quantities, in the measurement process. In this work we address this issue by  considering the measurement process that conserves an additive quantity across the object system that is measured, $\s$, and the measuring apparatus, $\aa$, and  derive  sufficient conditions of the measurement process so that decoherence does not affect the result.  First, we show that if the observable $O\sub{\s}$ commutes with the object system component of the conserved quantity $L\sub{\s}$, then  the conditional expectation value of $O\sub{\s}$ will not be sensitive to the coherence in the object system with respect to $L\sub{\s}$ so long as the apparatus state also commutes with the apparatus conserved quantity $L\sub{\aa}$.  Second, we show that if the measurement process is generated by a  Jaynes-Cummings Hamiltonian (which conserves the total excitation number) then the conditional expectation value of an observable that commutes with the number operator will be insensitive to the initial coherence in the system, even if the apparatus state does not commute with the number operator, so long as the compound system is symmetric in the number representation. This situation is relevant for experiments tracking quantum trajectories for qubit in resonant cavities and we comment on the consequence of our result for thermodynamic quantities investigated in these setups.

\section{Measurement models and conservation laws}
\subsection{Quantum measurements}
An observable on a quantum system $\s$, with Hilbert space $\h\sub{\s}$, is described by a positive operator valued measure (POVM) $M := \{M(x)\}_{x \in \xx}$, where $\xx$ denotes the set of measurement outcomes  and $M(x)$ are  positive \emph{effect} operators acting on $\h\sub{\s}$ that sum to the identity  \cite{PaulBuschMarianGrabowski1995}. Given an initial preparation of the system in state $\rho$, the probability of observing outcome $x$ of the POVM $M$ is given by the Born rule as 
\begin{align}
p_\rho^M(x) := \tr[M(x) \rho].
\end{align}
The physical implementation of a  POVM, by means of a suitable interaction with an apparatus, is referred to as a \emph{measurement model} \cite{Busch1996, Busch2016a}. Every POVM $M$ admits infinitely many measurement models, described by the tuple  $\mm := (\h\sub{\aa}, \varrho, U, Z\sub{\aa})$. Here $\h\sub{\aa}$ is the Hilbert space of apparatus $\aa$, and $\varrho$ is the state in which the apparatus is initially prepared; $U$ is a  unitary operator acting on $\h\sub{\s}\otimes \h\sub{\aa}$; and $Z\sub{\aa} = \sum_{x\in \xx} x P\sub{\aa}^x$ is a self-adjoint operator defining a projective valued measure (PVM) on $\aa$, where $P\sub{\aa}^x$ denote projection operators on $\h\sub{\aa}$.  Every outcome $x$  of the PVM $Z\sub{\aa}$ on $\aa$ is associated with outcome $x$ of the POVM $M$ on $\s$. Consequently, a measurement model can be considered as a method of transferring information from the object system $\s$ to the measurement apparatus $\aa$ by the unitary operator $U$, such that a  measurement of the apparatus by $Z\sub{\aa}$ after this process will replicate the statistics of directly measuring the object system by the POVM $M$. Because of this, the unitary interaction stage of measurement is referred to as ``premeasurement'', while the projective measurement at the end, responsible for leaving a permanent record of measurement outcomes, is referred to as ``objectification'' \cite{PeterMittelstaedt2004}. 

 For each measurement outcome $x$, the measurement model defines an instrument \cite{Heinosaari2011} on $\s$, given as
\begin{align}\label{eq:instrument-outcome}
\ii_x ^\mm (\rho) := \tr\sub{\aa}[(\one\sub{\s}\otimes P\sub{\aa}^x) U(\rho \otimes \varrho) U^\dagger], 
\end{align}
where $\tr\sub{\aa}[\cdot]$ denotes the partial trace over $\h\sub{\aa}$. The probability reproducibility criterion, which ensures that $\mm$ replicates the measurement statistics of $M$, is defined as 
\begin{align}
p_\rho^M(x) = \tr[\ii_x^\mm(\rho)] \text{ for all } \rho \text{ on } \h\sub{\s}.
\end{align}

\subsection{Expected value of a self-adjoint operator conditioned on the outcome of a POVM}

The state of the system, after observing outcome $x$ of the POVM $M$, implemented by the measurement model $\mm$, is denoted as $\rho(x) := \ii_x^\mm(\rho)/p_\rho^M(x)$. The conditional expectation value of a self adjoint operator $O\sub{\s}$, evaluated \emph{after} observing outcome $x$ of the POVM $M$, given an initial preparation of the system in state $\rho$, is thus given as 
\begin{align}\label{eq:O-after}
\<O\sub{\s}\>_\mathrm{after}^{\rho, \mm, x} &:= \tr[O\sub{\s} \rho(x) ], \nonumber \\
& = \frac{1}{p_\rho^M(x)} \tr[(O\sub{\s}\otimes P\sub{\aa}^x) U(\rho \otimes \varrho) U^\dagger].
\end{align}
The average value of $\<O\sub{\s}\>_\mathrm{after}^{\rho, \mm, x}$, over all measurement outcomes, is simply
\begin{align}
\sum_{x \in \xx} p_\rho^M(x)\<O\sub{\s}\>_\mathrm{after}^{\rho, \mm, x} = \tr[(O\sub{\s} \otimes \one\sub{\aa}) U(\rho \otimes \varrho) U^\dagger].
\end{align}

Similarly, we may define the conditional expectation value of $O\sub{\s}$, evaluated \emph{before} observing outcome $x$ of the POVM $M$, given an initial preparation of the system in state $\rho$,  as the real part of the generalised weak value of $O\sub{\s}$ \cite{Haapasalo2011}
\begin{align}\label{eq:O-before}
\<O\sub{\s}\>_\mathrm{before}^{\rho, M, x} &:= \frac{\mathrm{Re}\left(\tr[M(x) O\sub{\s} \rho]\right)}{p_\rho^M(x)} \equiv \frac{\mathrm{Re}\left(\tr[\ii_x^\mm( O\sub{\s} \rho)]\right)}{p_\rho^M(x)}, \nonumber \\
& = \frac{1}{2 p_\rho^M(x)} \tr[(\one\sub{\s}\otimes P\sub{\aa}^x) U((O\sub{\s}\rho + \rho O\sub{\s}) \otimes \varrho) U^\dagger],
\end{align}
while the average value of $\<O\sub{\s}\>_\mathrm{before}^{\rho,M, x}$ over all measurement outcomes is 
\begin{align}
\sum_{x \in \xx} p_\rho^M(x)\<O\sub{\s}\>_\mathrm{before}^{\rho, M, x} = \tr[O\sub{\s} \rho].
\end{align}
Note that while    while  $\<O\sub{\s}\>_\mathrm{after}^{\rho, \mm, x}$ depends on the specific measurement model $\mm$ for the POVM $M$, the same is not true for $\<O\sub{\s}\>_\mathrm{before}^{\rho, M, x}$, which  is uniquely determined by the POVM $M$.  Finally, we may define the conditional change in the quantity $O\sub{\s}$, given outcome $x$ of the POVM $M$, as the difference between \eq{eq:O-before} and \eq{eq:O-after}, which is 
\begin{align}\label{eq:O-change}
\Delta O\sub{\s}^{\rho, \mm, x} := \<O\sub{\s}\>_\mathrm{after}^{\rho, \mm, x} - \<O\sub{\s}\>_\mathrm{before}^{\rho, M, x}. 
\end{align}
The average change in this quantity is thus simply $\<\Delta O\sub{\s}^{\rho, \mm, x}\> =\tr[(O\sub{\s} \otimes \one\sub{\aa}) U(\rho \otimes \varrho) U^\dagger] - \tr[O\sub{\s} \rho] $.

While \eq{eq:O-after} and \eq{eq:O-before} can only be interpreted as statistical properties of the system in general,  they can be definite properties under specific conditions. The most trivial case is for \eq{eq:O-after}, which is a definite property if   $\rho(x)$ only has support on a single (possibly degenerate) subspace of $O\sub{\s}$.  The more interesting case where the conditional expectation value can be a definite property is for \eq{eq:O-before}. In the special case where $\rho$ is a pure state $\ket{\psi}$, and $M(x)$ is a projection on the pure state $\ket{\phi}$, \eq{eq:O-before} simplifies to the real component of the familiar weak value $\<\phi|O\sub{\s}|\psi\>/\<\phi|\psi\>$. As discussed in \cite{weak-value-beyond-exp}, in such a case the weak value can be interpreted as the ``eigenvalue'' obtained by measuring the observable $O\sub{\s}$  on the system described by the two-state vector \cite{weak-value-two-time}  $\<\phi| \,  |\psi\>$, where $\ket{\psi}$ describes the evolution of the system forwards in time, while  $\ket{\phi}$ describes the evolution of the system backwards in time.  To see this, consider the case where the measurement model for the  observable $O\sub{\s}$ is given by an apparatus that is a particle on a line, given by a pure state described by the position operator $Q$, and the premeasurement unitary interaction with the system is $U(g) = e^{-i g O\sub{\s} \otimes P}$, with $P$ the conjugate momentum to $Q$, and $g$ a strength parameter.  If $\ket{\psi}$ is an eigenstate of $O\sub{\s}$, with eigenvalue $o$, the apparatus will remain in a pure state after the measurement interaction, and its position will shift by $go$. As such, the system in state $\ket{\psi}$ has a definite property of the observable $O\sub{\s}$. If, however, $\ket{\psi}$ is not an eigenstate of $O\sub{\s}$, then the apparatus will be in a statistical mixture of pure states, each shifted by a different amount; here, $\ket{\psi}$ will not have a definite property of $O\sub{\s}$. But if we also post-select the system onto the pure state $\ket{\phi}$ after its interaction with the apparatus, then if $g$ is sufficiently small the apparatus will remain pure, with its position shifted by $\approx g \<\phi|O\sub{\s}|\psi\>/\<\phi|\psi\>$; analogously, we may say that the two-state vector $\<\phi| \,  |\psi\>$ has a definite property of the observable $O\sub{\s}$, even though this property may lie outside the range of the eigenvalues of $O\sub{\s}$. However,  such an interpretation does not hold in the general case,  where either the initial state $\rho$ is mixed, or the POVM element $M(x)$ is not a projection on a pure state. This is because in such a case, the apparatus will again be in a statistical mixture of pure states, each shifted by a different amount.

\subsection{Measurements restricted by additive conservation laws}
Measurements, as other physical processes, may be subject to conservation laws. These are characterised by the commutation of the measurement unitary operator $U$ with some quantity $L$.  A class of interest are additive conservation laws,  meaning that $L = L\sub{\s} + L\sub{\aa}$, with $L\sub{\s}$ and $L\sub{\aa}$ being self adjoint operators on $\h\sub{\s}$ and $\h\sub{\aa}$, respectively.  

In the presence of additive conservation laws, measurements will be restricted by the  Wigner-Araki-Yanase (WAY) theorem \cite{E.Wigner1952,Araki1960,Loveridge2011}. Consider a measurement model $\mm := (\h\sub{\aa}, \varrho, U, Z\sub{\aa})$ that defines a PVM $M$ on the system $\s$, such that $U$ commutes with $L\sub{\s} + L\sub{\aa}$. The WAY theorem states that either if $M$ is repeatable, or $Z\sub{\aa}$ commutes with $L\sub{\aa}$ (the Yanase condition), it will follow that $M$ must commute with $L\sub{\s}$. 

A PVM $M$ is said to be repeatable if, conditional on observing outcome $x$, a subsequent measurement of $M$ will result in outcome $x$ with certainty. While it is not necessary for a measurement of $M$ on $\s$ to be repeatable, it is necessary for the measurement of $Z\sub{\aa}$ on the apparatus to be repeatable; this is because the record of a measurement outcome stored in the apparatus must be a permanent fixture of the world. 

Measurement models can be extended ad infinitum, by means of introducing a measurement model for the apparatus observable $Z\sub{\aa}$ with the aid of an additional apparatus $\bb$. Therefore, if the measurement of $Z\sub{\aa}$ is to be implemented under an additive conservation law $L\sub{\aa} + L\sub{\bb}$, then by the WAY theorem the requirement of repeatability will necessitate that $Z\sub{\aa}$ commutes with $L\sub{\aa}$.   

\section{Results}

\subsection{Conserved quantities and decoherence maps}
 Let us now define the decoherence map with respect to a self-adjoint operator $L$ as 
\begin{align}\label{eq:decoherence-map}
\Phi_L(A) := \sum_l Q^l A Q^l,
\end{align}
where $Q^l$ are the spectral projections of $L$. If any physical phenomenon, such as measurement, does not distinguish between $A$ and $\Phi_L(A)$, i.e., the two states give the same result, we may say that the coherence of $A$ with respect to $L$ is a symmetry of that phenomenon. 

We now explore the situations in which the coherence of the system and apparatus states $\rho$ and $\varrho$, with respect to the conserved quantities $L\sub{\s}$ and $L\sub{\aa}$, will be symmetries of  $\Delta O\sub{\s}^{\rho, \mm, x} $, defined in \eq{eq:O-change}, if $O\sub{\s}$ commutes with $L\sub{\s}$. First, we shall prove that if a measurement model conserves an additive quantity $L= L\sub{\s} + L\sub{\aa}$, then the coherence of the object system state $\rho$ (apparatus system state $\varrho$) with respect to $L\sub{\s}$ ($L\sub{\aa}$) will be a symmetry of $\Delta O\sub{\s}^{\rho, \mm, x}$, if $O\sub{\s}$ commutes with $L\sub{\s}$ and $\varrho$ commutes with $L\sub{\aa}$ ($\rho$ commutes with $L\sub{\s}$).

\begin{thm}\label{thm:Theorem 1}
Consider the POVMs $M_1$ and $M_2$, induced by the measurement models $\mm_1:= (\h\sub{\aa}, \varrho_1, U, Z\sub{\aa})$ and $\mm_2:= (\h\sub{\aa}, \varrho_2, U, Z\sub{\aa})$.  Let $U$ commute with $L = L\sub{\s}+ L\sub{\aa}$ and $Z\sub{\aa}$ commute with $L\sub{\aa}$.  Finally, let $\varrho_2 = \Phi_{L\sub{\aa}}(\varrho_1)$, with $\Phi_{L\sub{\aa}}$ defined by \eq{eq:decoherence-map}.  It follows that if both $O\sub{\s}$  and $\rho$ commute with $L\sub{\s}$, then 
\begin{align}\label{eq:system_commutes}
\<O\sub{\s}\>_\mathrm{before}^{\rho, M_1, x} &= \<O\sub{\s}\>_\mathrm{before}^{\rho, M_2, x}, \nonumber \\
\<O\sub{\s}\>_\mathrm{after}^{\rho, \mm_1, x} &= \<O\sub{\s}\>_\mathrm{after}^{\rho, \mm_2, x}.
\end{align}
Conversely, if  $O\sub{\s}$ commutes with $L\sub{\s}$, then for any $\rho$
\begin{align}\label{eq:ancilla_commutes}
\<O\sub{\s}\>_\mathrm{before}^{\rho, M_2, x} &= \<O\sub{\s}\>_\mathrm{before}^{\Phi_{L\sub{\s}}(\rho), M_2, x}, \nonumber \\
\<O\sub{\s}\>_\mathrm{after}^{\rho, \mm_2, x} &= \<O\sub{\s}\>_\mathrm{after}^{\Phi_{L\sub{\s}}(\rho), \mm_2, x}.
\end{align}
\end{thm}
The proof is provided in \app{app:proofs}. \eq{eq:system_commutes} shows that the coherence in the apparatus state with respect to the conserved quantity $L\sub{\aa}$ is a symmetry of $\Delta O\sub{\s}^{\rho,\mm,x}$ if both $O\sub{\s}$ and $\rho$ commute with the conserved quantity $L\sub{\s}$. Meanwhile,  \eq{eq:ancilla_commutes} shows that the coherence of the system state with respect to the conserved quantity $L\sub{\s}$ is a symmetry of $\Delta O\sub{\s}^{\rho,\mm,x}$ if $O\sub{\s}$ commutes with $L\sub{\s}$ and the apparatus state $\varrho$ commutes with $L\sub{\aa}$. 

It is, however, possible for the coherence in both $\rho$ and $\varrho$ to be a symmetry of $\Delta O\sub{\s}^{\rho, \mm, x}$ even if neither of them commutes with the conserved quantity. 

\begin{thm}\label{thm:Theorem 2}
Consider the POVMs $M_1$ and $M_2$, induced by the measurement models $\mm_1:= (\h\sub{\aa}, \varrho_1, U, Z\sub{\aa})$ and $\mm_2:= (\h\sub{\aa}, \varrho_2, U, Z\sub{\aa})$.  Let $U$ commute with $L = L\sub{\s}+ L\sub{\aa}$ and $Z\sub{\aa}$ commute with $L\sub{\aa}$. Finally, let $\varrho_2 = \Phi_{L\sub{\aa}}(\varrho_1)$, with $\Phi_{L\sub{\aa}}$ defined by \eq{eq:decoherence-map}. Denote the eigenvectors of $L\sub{\s}$ and $L\sub{\aa}$ as $\ket{\phi_m^\alpha}$ and $\ket{\varphi_\mu^\beta}$ respectively, where $m$ and $\mu$ are eigenvalues, while $\alpha$ and $\beta$ label the degeneracy. It follows that if $O\sub{\s}$ commutes with $L\sub{\s}$, $\rho\otimes \varrho_i$ is symmetric in the eigenbasis representation $\ket{\phi_m^\alpha \otimes \varphi_\mu^\beta}$, and the real component of $\<\phi_n^{\alpha'}\otimes \varphi_\nu^{\beta'}|U^\dagger(O\sub{\s} \otimes P\sub{\aa}^x ) U|\phi_m^{\alpha}\otimes \varphi_\mu^{\beta}\> $ is zero when $m \ne n$ and $\mu \ne \nu$, then 
\begin{align}\label{eq:exp-O-symmetric}
\<O\sub{\s}\>_\mathrm{before}^{\rho, M_1, x} & = \<O\sub{\s}\>_\mathrm{before}^{\rho, M_2, x} = \<O\sub{\s}\>_\mathrm{before}^{\Phi_{L\sub{\s}}(\rho), M_1, x}  = \<O\sub{\s}\>_\mathrm{before}^{\Phi_{L\sub{\s}}(\rho), M_2, x}, \nonumber \\
\<O\sub{\s}\>_\mathrm{after}^{\rho, \mm_1, x}&= \<O\sub{\s}\>_\mathrm{after}^{\rho, \mm_2, x} = \<O\sub{\s}\>_\mathrm{after}^{\Phi_{L\sub{\s}}(\rho), \mm_1, x}= \<O\sub{\s}\>_\mathrm{after}^{\Phi_{L\sub{\s}}(\rho), \mm_2, x}.
\end{align}
\end{thm}
The proof is provided in \app{app:proofs}. It will be instructive to consider physical situations where \thmref{thm:Theorem 2} applies.

\subsection{Qubits measured by a Jaynes-Cummings interaction}

A simple case of measurement process with an additive  conserved quantity is obtained by a Jaynes-Cummings system-apparatus interaction.  Consider the case where the measurement unitary operator is $U = e^{-i \theta H_I}$, with the  Jaynes-Cummings interaction Hamiltonian
\begin{align}\label{eq:Jaynes-Cummings}
H_I := \sigma\sub{\s}^+ \otimes \sigma\sub{\aa}^- +  \sigma\sub{\s}^- \otimes \sigma\sub{\aa}^+
\end{align}
where for $X \in \{\s,\aa\}$, 
\begin{align}
\sigma\sub{X}^+:= \sum_{k=0}^{d\sub{X} -1}\sqrt{k+1}|k+1\>\<k| = (\sigma\sub{X}^-)^\dagger.
\end{align}
Here, $d\sub{X}$ is the dimension of Hilbert space $\h\sub{X}$, and $\ket{k}$ denotes the excitation number $k$ of the system. The unitary $U$ conserves the total excitation number $\nn = \nn\sub{\s} + \nn\sub{\aa}$, where for $X \in \{\s,\aa\}$,
\begin{align}\label{eq:number operator}
\nn\sub{X} = \sum_{k=0}^{d\sub{X} - 1}k |k\>\<k|. 
\end{align}
Given the observables $O\sub{\s}$ and $Z\sub{\aa}$ that commute with $\nn\sub{\s}$ and $\nn\sub{\aa}$ respectively, it follows that the real component of $\<n\otimes\nu|U^\dagger (O\sub{\s} \otimes P\sub{\aa}^x) U|m\otimes \mu\>$ will always be zero when $n\ne m$ and $\nu \ne \mu$ so long as either $d\sub{\s}=2$ or $d\sub{\aa}=2$. Let us consider the case where $d\sub{\s}=2$, i.e., when the object system being measured is a qubit, while the measuring apparatus can have an arbitrarily large dimension.   It follows that $U = \sum_{l=0}^{d\sub{\aa}+1} U_l$, where for $1\leqslant l\leqslant d\sub{\aa}$, $U_l$ is a 2-dimensional matrix acting on the subspace spanned by $\{\ket{0\otimes l}, \ket{1\otimes l-1}\}$, given as $U_l = e^{-i \theta \sqrt{l} \sigma_x} = \cos(\theta \sqrt{l})\one -i \sin(\theta \sqrt{l})\sigma_x$, with $\sigma_x$ the Pauli-X matrix defined as $\sigma_x \ket{0\otimes l} = \ket{1\otimes l-1}$. Consequently,  for $n\ne m$ and $\nu \ne \mu$, the real part of $\<n\otimes \nu|U|m\otimes \mu\>$ is zero, while the imaginary part of $\<m\otimes \mu|U|m\otimes \mu\>$ is zero. Consequently, by choosing $O\sub{\s} = \lambda_0 |0\>\<0| + \lambda_1 |1\>\<1|$ we may write
\begin{align}
\<0\otimes l|U^\dagger (O\sub{\s} \otimes P\sub{\aa}^x) U|1\otimes l-1\> &= \lambda_0 \delta_{l,x} \<0\otimes l|U^\dagger |0 \otimes l\> \<0 \otimes l|U|1 \otimes l-1\> \nonumber \\
&\qquad + \lambda_1 \delta_{l-1,x} \<0\otimes l|U^\dagger |1 \otimes l-1\> \<1 \otimes l-1|U|1 \otimes l-1\>,
\end{align}
where $\delta_{k,x}$ is the Kronecker delta function determining if $\ket{k}$ lies in the range of the projector $P\sub{\aa}^x$. Each term of the above equation is a product of a purely real number, with a purely imaginary number. Therefore, the total value is purely imaginary. 

\begin{figure*}
\includegraphics[width = 0.6\textwidth]{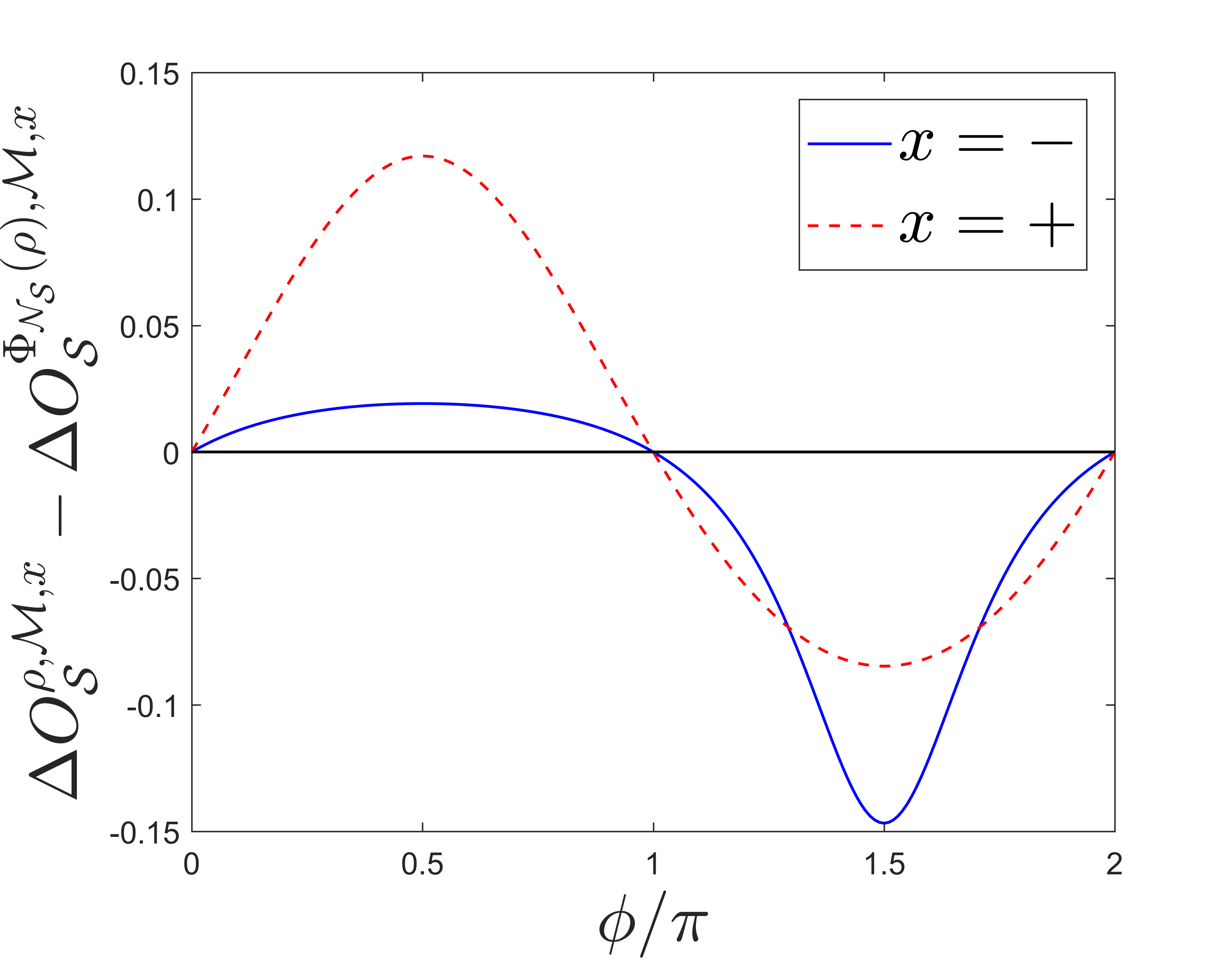} 
\caption{The measurement model for the POVM on $\h\sub{\s}$ is $\mm:= (\h\sub{\aa}, \ket{\xi}, U, Z\sub{\aa})$. Here,  the measurement unitary is given as $U = e^{-i (\pi/3) H_I}$, with the Jaynes Communigs interaction Hamiltonian defined in \eq{eq:Jaynes-Cummings}. This conserves the total excitation number $\nn\sub{\s} + \nn\sub{\aa}$, defined in \eq{eq:number operator}. The apparatus observable is $Z\sub{\aa} = |1\>\<1| - |0\>\<0|$, with the corresponding outcomes $x =\pm$. The initial state of the apparatus is the coherent state $\ket{\xi}= \cos(\pi/6)|1\> + \sin(\pi/6) |0\>$, while the object system is  initially prepared in the coherent state $\ket{\psi} = \cos(\pi/8)|1\> +e^{i \phi} \sin(\pi/8) |0\>$. Finally, the system observable is $O\sub{\s} = |1\>\<1| - |0\>\<0|$. When $\phi \in \{0, \pi\}$, the state $|\psi\>\<\psi|\otimes |\xi\>\<\xi|$ is symmetric in the excitation number representation $|m,\mu\>$, and so by \thmref{thm:Theorem 2}  $\Delta O\sub{\s}^{\rho, \mm, x} = \Delta O\sub{\s}^{\Phi_{\nn\sub{\s}}(\rho), \mm, x}$, with these quantities defined by \eq{eq:O-change}. }   \label{fig:Delta_O_rho_phi}
\end{figure*}

Given such a model, the results of \thmref{thm:Theorem 2} will follow so long as the initial product state of system and apparatus, $\rho\otimes \varrho$, is symmetric in the excitation number representation. As such, let us consider the simplest example where both $\s$ and $\aa$ are qubits, where $\rho = |\psi\>\<\psi|$, with $\ket{\psi}:= \cos(\theta_1/2) \ket{1} + e^{i \phi}\sin(\theta_1/2) \ket{0}$, while $\varrho = |\xi\>\<\xi|$, with $\ket{\xi} := \cos(\theta_2/2)\ket{1} + \sin(\theta_2/2) \ket{0}$. Moreover, let us choose $Z\sub{\aa} = |1\>\<1| - |0\>\<0|$ (with outcomes $x=\pm$) and $O\sub{\s} = |1\>\<1| - |0\>\<0|$. According to \thmref{thm:Theorem 2}, therefore, $\Delta O\sub{\s}^{\rho, \mm, x} = \Delta O\sub{\s}^{\Phi_{\nn\sub{\s}}(\rho), \mm, x}$, with these quantities defined in \eq{eq:O-change}, if $\phi \in \{0, \pi\}$. This is shown in \fig{fig:Delta_O_rho_phi}, where $\Delta O\sub{\s}^{\rho, \mm, x} - \Delta O\sub{\s}^{\Phi_{\nn\sub{\s}}(\rho), \mm, x}$ is plotted as a function of $\phi$, and deviates from zero as soon as $\phi \ne 0, \pi$.

\section{Discussion}
Our findings have physical implications in all cases where conditional measurements are associated with physical quantities. A striking example is that of energy changes along quantum trajectories \cite{Mohammady2018a}, with its thermodynamic implications. In particular, \thmref{thm:Theorem 1} can be seen as an extension of results pertaining to thermal operations to thermal measurements. Recall that a thermal operation is constituted of an energy conserving unitary interaction with an apparatus that is prepared in a Gibbs state. It is known that thermal operations are a subset of time-translation symmetric operations \cite{Lostaglio2015b}. This implies that, since $U$ commutes with the total Hamiltonian $H\sub{\s}+ H\sub{\aa}$, and the apparatus state $\varrho$ commutes with $H\sub{\aa}$ (which is clearly the case when $\varrho$ is a Gibbs state $e^{-H\sub{\aa}/k_B T}/ \tr[e^{-H\sub{\aa}/k_B T}]$), then  $\tr[(H\sub{\s}\otimes \one\sub{\aa})U(\rho\otimes \varrho)U^\dagger] = \tr[(H\sub{\s}\otimes \one\sub{\aa})U(\Phi_{H\sub{\s}}(\rho)\otimes \varrho)U^\dagger]$. That is to say, the coherence in $\rho$ with respect to $H\sub{\s}$ is a symmetry of the average change in energy for thermal operations. A thermal measurement can be seen as augmenting a thermal operation with a projective measurement of the apparatus by some observable $Z\sub{\aa}$. However, conditional energy changes given a thermal measurement are not necessarily invariant with respect to the coherence in the system;  as shown in \thmref{thm:Theorem 1}, for the conditional change in energies to be invariant with respect to the coherence in $\rho$,  the Yanase condition must satisfied, i.e.,  the apparatus observable $Z\sub{\aa}$ must  commute with $H\sub{\aa}$. 

In \thmref{thm:Theorem 2} we show how, in some circumstances, the coherence in the system does not affect the conditional change in expectation values even if  the apparatus state $\varrho$ does not commute with its conserved quantity.   A common example of such a situation is where the object system being measured is a qubit, and the measurement interaction is generated by a Jaynes-Cummings Hamiltonian, with the  conserved quantity being the total excitation number. This setup describes, for example, the coupling of a superconducting qubit and a cavity mode used to track the energy changes of the system along its trajectories [Kater].  We show that so long as the state of the compound system prior to measurement is symmetric in the number representation, then the coherence in the object system w.r.t the number operator will not affect conditional expectation values of observables that commute with the number operator, such as the energy.

\section{Conclusions}

In this work we have studied the effect of  quantum coherence on the expectation values of observables, conditioned on the outcome of generalized measurements, subject to symmetry constraints.  To this end we have considered a general measurement model which preserves a quantity additive in the system and apparatus degrees of freedom (additive conservation law). We then  identified  the sufficient conditions  such that the conditional expectation values given some initial state $\rho$ will be identical to that given when the coherence of the initial state is removed.   We have finally illustrated our results for a simple measurement model of a qubit coupled to a harmonic oscillator (e.g. an atom in a resonant electromagnetic cavity). The conditional expectation value can be controlled by tuning a relative phase of the qubit state and it is generically different from its incoherent counterpart; the two being equal for the value of the phase that fulfil the theorem conditions. This shows that additive symmetries, present in systems used for quantum measurement experiments, can strongly constrain the role of system coherences, to the point that they might not affect the conditional expectation values of observables.

\acknowledgments

This research was supported by EPSRC (Grant EP/P030815/1)

%%%%%%%%%%

%\newpage

%\medskip
\bibliographystyle{apsrev4-1}
\bibliography{Thermodynamics-Coherence-Measurement}

%\newpage

%\onecolumngrid
%\begin{widetwxt}

%\setcounter{equation}{0}
%\setcounter{figure}{0}
%\setcounter{table}{0}
%\setcounter{page}{1}
%\makeatletter
%\renewcommand{\theequation}{S\arabic{equation}}
%\renewcommand{\thefigure}{S\arabic{figure}}
%\renewcommand{\bibnumfmt}[1]{[S#1]}
%\renewcommand{\citenumfont}[1]{S#1}

\makeatletter
\renewcommand\p@subsection{\thesection\,}
\makeatother
\makeatletter
\renewcommand\p@subsubsection{\thesection\,\thesubsection\,}
\makeatother

\appendix

\section{Proofs of Theorems}\label{app:proofs}
\begin{proof}[Proof of Theorem 1]
Let us denote the spectral projections of $L$ as $Q^l = \sum_{(m,\mu)_l}Q\sub{\s}^m \otimes Q\sub{\aa}^\mu$, where $Q\sub{\s}^m$ and $Q\sub{\aa}^\mu$ are the spectral projections of $L\sub{\s}$ and $L\sub{\aa}$ respectively, and $(m,\mu)_l$ denotes the pair of eigenvalues which sum to  $m+\mu = l$. The commutation relation $[U, L]_-=\zero$ will therefore imply that $U = \sum_l Q^l U Q^l $, and so we have
\begin{align}
\<O\sub{\s}\>_\mathrm{after}^{\rho, \mm_i, x} &:=  \frac{1}{p_\rho^{M_i}(x)} \tr[(O\sub{\s}\otimes P\sub{\aa}^x) U(\rho \otimes \varrho_i) U^\dagger], \nonumber \\
&=\frac{1}{p_\rho^{M_i}(x)} \sum_{l,l'}\tr[(O\sub{\s}\otimes P\sub{\aa}^x)Q^l U Q^{l}(\rho \otimes \varrho_i) Q^{l'} U^\dagger Q^{l'}], \nonumber \\
&= \frac{1}{p_\rho^{M_i}(x)} \sum_{l,l'}\sum_{(m,\mu)_l}\sum_{(n,\nu)_{l'}} \tr[(Q\sub{\s}^n O\sub{\s}Q\sub{\s}^m \otimes Q\sub{\aa}^\nu P\sub{\aa}^x Q\sub{\aa}^\mu) U Q^{l}(\rho  \otimes  \varrho_i )Q^{l'} U^\dagger].
\end{align}
Since $Z\sub{\aa}$  commutes with $L\sub{\aa}$, and $O\sub{\s}$ commutes with $L\sub{\s}$, the last line of this equation may be rewritten as 
\begin{align}
\<O\sub{\s}\>_\mathrm{after}^{\rho, \mm_i, x} &= \frac{1}{p_\rho^{M_i}(x)} \sum_{l,l'}\sum_{(m,\mu)_l}\sum_{(m,\mu)_{l'}} \tr[(Q\sub{\s}^m O\sub{\s}Q\sub{\s}^m \otimes Q\sub{\aa}^\mu P\sub{\aa}^x Q\sub{\aa}^\mu) U Q^{l}(\rho  \otimes  \varrho_i )Q^{l'} U^\dagger], \nonumber \\
&= \frac{1}{p_\rho^{M_i}(x)} \sum_{l}\tr[(O\sub{\s}\otimes P\sub{\aa}^x) U Q^{l}(\rho \otimes \varrho_i) Q^{l} U^\dagger ].
\end{align}
Expanding $Q^l$ into the spectral projections of $L\sub{\s}$ and $L\sub{\aa}$ will thus yield the expression 
\begin{align}\label{eq:summetry-O-after}
\<O\sub{\s}\>_\mathrm{after}^{\rho, \mm_i, x} &= \frac{1}{p_\rho^{M_i}(x)} \sum_{l}\sum_{(m,\mu)_l}\sum_{(n,\nu)_{l}} \tr[(O\sub{\s} \otimes P\sub{\aa}^x ) U (Q\sub{\s}^m\rho Q\sub{\s}^n  \otimes Q\sub{\aa}^\mu \varrho_i Q\sub{\aa}^\nu) U^\dagger].
\end{align}
Similarly, we may write 
\begin{align}\label{eq:summetry-O-before}
\<O\sub{\s}\>_\mathrm{before}^{\rho, M_i, x}  &= \frac{1}{2 p_\rho^{M_i}(x)} \sum_{l}\sum_{(m,\mu)_l}\sum_{(n,\nu)_{l}} \tr[(\one\sub{\s} \otimes P\sub{\aa}^x ) U ( Q\sub{\s}^m (O\sub{\s}  \rho+ \rho O\sub{\s}) Q\sub{\s}^n  \otimes Q\sub{\aa}^\mu \varrho_i Q\sub{\aa}^\nu) U^\dagger].
\end{align}
If $\rho$ commutes with $L\sub{\s}$, then the only terms that remain in  \eq{eq:summetry-O-after} and \eq{eq:summetry-O-before} are those with $m=n$. The fact that $m + \mu = m + \nu = l$ thus implies that $\mu = \nu$, and so we may make the substituion $\varrho_i = \Phi_{L\sub{\aa}}(\varrho_i)$, which gives  \eq{eq:system_commutes}. Conversely, for the POVM $M_2$, $\mu = \nu$, and by the same argument $m=n$, which implies \eq{eq:ancilla_commutes}.
\end{proof}

\begin{proof}[Proof of Theorem 2]
Recall from \thmref{thm:Theorem 1} that, given the commutation relations $[U,L\sub{\s} + L\sub{\aa}]_-=\zero$, $[O\sub{\s},L\sub{\s} ]_-=\zero$, and $[Z\sub{\aa}, L\sub{\aa}]_-=\zero$, we arrive at \eq{eq:summetry-O-after} and \eq{eq:summetry-O-before}. We may write \eq{eq:summetry-O-after} as 
\begin{align}\label{eq:O-after-expansion}
\<O\sub{\s}\>_\mathrm{after}^{\rho, \mm_i, x} 
& = \frac{1}{p_\rho^{M_i}(x)} \sum_{l}\sum_{(m,\mu)_l} \tr[(O\sub{\s} \otimes P\sub{\aa}^x ) U (Q\sub{\s}^m\rho Q\sub{\s}^m  \otimes Q\sub{\aa}^\mu \varrho_i Q\sub{\aa}^\mu) U^\dagger] \nonumber \\
& + \frac{1}{p_\rho^{M_i}(x)}  \sum_{l}\sum_{(m,\mu)_l}\sum_{(n\ne m,\nu\ne \mu)_{l}} \tr[(O\sub{\s} \otimes P\sub{\aa}^x ) U (Q\sub{\s}^m\rho Q\sub{\s}^n  \otimes Q\sub{\aa}^\mu \varrho_i Q\sub{\aa}^\nu) U^\dagger].
\end{align}
Similarly, we may write \eq{eq:summetry-O-before} as 
\begin{align}\label{eq:O-before-expansion}
\<O\sub{\s}\>_\mathrm{before}^{\rho, M_i, x} & = \frac{1}{p_\rho^{M_i}(x)} \sum_{l}\sum_{(m,\mu)_l}\left( \tr[(\one\sub{\s} \otimes P\sub{\aa}^x ) U ( O\sub{\s} Q\sub{\s}^m \rho Q\sub{\s}^m  \otimes Q\sub{\aa}^\mu \varrho_i Q\sub{\aa}^\mu) U^\dagger]\right) \nonumber \\
&+ \frac{1}{2p_\rho^{M_i}(x)}  \sum_{l}\sum_{(m,\mu)_l}\sum_{(n\ne m,\nu\ne \mu)_{l}}\left( \tr[(\one\sub{\s} \otimes P\sub{\aa}^x ) U (  Q\sub{\s}^m (O\sub{\s}\rho + \rho O\sub{\s}) Q\sub{\s}^n  \otimes Q\sub{\aa}^\mu \varrho_i Q\sub{\aa}^\nu) U^\dagger]\right).
\end{align}
For a given $l$,  $m \ne n$, and $\mu \ne \nu$ the terms in \eq{eq:O-after-expansion} can be expanded as
\begin{align}\label{eq:O-after-expansion-mnen}
&\tr[(O\sub{\s} \otimes P\sub{\aa}^x ) U (Q\sub{\s}^m\rho Q\sub{\s}^n  \otimes Q\sub{\aa}^\mu \varrho_i Q\sub{\aa}^\nu) U^\dagger] + \tr[(O\sub{\s} \otimes P\sub{\aa}^x ) U (Q\sub{\s}^n\rho Q\sub{\s}^m  \otimes Q\sub{\aa}^\nu \varrho_i Q\sub{\aa}^\mu) U^\dagger] \nonumber \\
& \qquad = \sum_{\alpha, \alpha', \beta, \beta'} \<\phi_n^{\alpha'}\otimes \varphi_\nu^{\beta'}|U^\dagger(O\sub{\s} \otimes P\sub{\aa}^x ) U|\phi_m^{\alpha}\otimes \varphi_\mu^{\beta}\>\<\phi_m^\alpha|\rho|\phi_n^{\alpha'}\>\<\varphi_\mu^\beta|\varrho_i|\varphi_\nu^{\beta'}\> \nonumber \\
& \qquad  + \sum_{\alpha, \alpha', \beta, \beta'} \<\phi_m^{\alpha}\otimes \varphi_\mu^{\beta} |U^\dagger(O\sub{\s} \otimes P\sub{\aa}^x ) U| \phi_n^{\alpha'}\otimes \varphi_\nu^{\beta'}\>\<\phi_n^{\alpha'}|\rho|\phi_m^{\alpha}\>\<\varphi_\nu^{\beta'}|\varrho_i|\varphi_\mu^{\beta}\>.
\end{align}
while those of \eq{eq:O-before-expansion} are expanded as 
\begin{align}\label{eq:O-before-expansion-mnen}
&\tr[(\one\sub{\s} \otimes P\sub{\aa}^x ) U (  Q\sub{\s}^m (O\sub{\s}\rho + \rho O\sub{\s}) Q\sub{\s}^n  \otimes Q\sub{\aa}^\mu \varrho_i Q\sub{\aa}^\nu) U^\dagger] + \tr[(\one\sub{\s} \otimes P\sub{\aa}^x ) U (  Q\sub{\s}^n (O\sub{\s}\rho + \rho O\sub{\s}) Q\sub{\s}^m  \otimes Q\sub{\aa}^\nu \varrho_i Q\sub{\aa}^\mu) U^\dagger] \nonumber \\
&\qquad = \sum_{\alpha, \alpha', \beta, \beta'} \<\phi_n^{\alpha'}\otimes \varphi_\nu^{\beta'}|U^\dagger(\one\sub{\s} \otimes P\sub{\aa}^x ) U|\phi_m^{\alpha}\otimes \varphi_\mu^{\beta}\>\<\phi_m^\alpha| (O\sub{\s}\rho + \rho O\sub{\s})|\phi_n^{\alpha'}\>\<\varphi_\mu^\beta|\varrho_i|\varphi_\nu^{\beta'}\> \nonumber \\
& \qquad  + \sum_{\alpha, \alpha', \beta, \beta'} \<\phi_m^{\alpha}\otimes \varphi_\mu^{\beta} |U^\dagger(\one\sub{\s} \otimes P\sub{\aa}^x ) U| \phi_n^{\alpha'}\otimes \varphi_\nu^{\beta'}\>\<\phi_n^{\alpha'}| (O\sub{\s}\rho + \rho O\sub{\s})|\phi_m^{\alpha}\>\<\varphi_\nu^{\beta'}|\varrho_i|\varphi_\mu^{\beta}\>.
\end{align}
If $\rho \otimes \varrho$ is symmetric in the representation of $\ket{\phi_m^{\alpha}\otimes \varphi_\mu^{\beta}}$, it follows that $\<\phi_m^\alpha|\rho|\phi_n^{\alpha'}\>\<\varphi_\mu^\beta|\varrho_i|\varphi_\nu^{\beta'}\> = \<\phi_n^{\alpha'}|\rho|\phi_m^{\alpha}\>\<\varphi_\nu^{\beta'}|\varrho_i|\varphi_\mu^{\beta}\>$, while $\<\phi_m^\alpha| (O\sub{\s}\rho + \rho O\sub{\s})|\phi_n^{\alpha'}\>\<\varphi_\mu^\beta|\varrho_i|\varphi_\nu^{\beta'}\> = \<\phi_n^{\alpha'}| (O\sub{\s}\rho + \rho O\sub{\s})|\phi_m^{\alpha}\> \<\varphi_\nu^{\beta'}|\varrho_i|\varphi_\mu^{\beta}\>$ and so we may write the right hand sides of \eq{eq:O-after-expansion-mnen} and \eq{eq:O-before-expansion-mnen} as
\begin{align}\label{eq:O-after-expansion-mnen-2}
\sum_{\alpha, \alpha', \beta, \beta'} &\left(\<\phi_n^{\alpha'}\otimes \varphi_\nu^{\beta'}|U^\dagger(O\sub{\s} \otimes P\sub{\aa}^x ) U|\phi_m^{\alpha}\otimes \varphi_\mu^{\beta}\> +  \text{complex conjugate} \right) \<\phi_m^\alpha|\rho|\phi_n^{\alpha'}\>\<\varphi_\mu^\beta|\varrho_i|\varphi_\nu^{\beta'}\>
\end{align}
and 
\begin{align}\label{eq:O-before-expansion-mnen-2}
\sum_{\alpha, \alpha', \beta, \beta'} &\left(\<\phi_n^{\alpha'}\otimes \varphi_\nu^{\beta'}|U^\dagger(\one\sub{\s} \otimes P\sub{\aa}^x ) U|\phi_m^{\alpha}\otimes \varphi_\mu^{\beta}\> +  \text{complex conjugate} \right) \<\phi_m^\alpha| (O\sub{\s}\rho + \rho O\sub{\s})|\phi_n^{\alpha'}\>\<\varphi_\mu^\beta|\varrho_i|\varphi_\nu^{\beta'}\>.
\end{align}
Finally, if $\<\phi_n^{\alpha'}\otimes \varphi_\nu^{\beta'}|U^\dagger(O\sub{\s} \otimes P\sub{\aa}^x ) U|\phi_m^{\alpha}\otimes \varphi_\mu^{\beta}\>$ is purely imaginary (which implies that so is $\<\phi_n^{\alpha'}\otimes \varphi_\nu^{\beta'}|U^\dagger(\one\sub{\s} \otimes P\sub{\aa}^x ) U|\phi_m^{\alpha}\otimes \varphi_\mu^{\beta}\>$), it follows that the terms inside the parantheses of \eq{eq:O-after-expansion-mnen-2} and \eq{eq:O-before-expansion-mnen-2} vanish. Consequently, the second lines of \eq{eq:O-after-expansion} and \eq{eq:O-before-expansion} will be zero, and we are left with 
\begin{align}
\<O\sub{\s}\>_\mathrm{after}^{\rho, \mm_i, x}  &= \frac{1}{p_\rho^M(x)} \sum_{l}\sum_{(m,\mu)_l} \tr[(O\sub{\s} \otimes P\sub{\aa}^x ) U (Q\sub{\s}^m\rho Q\sub{\s}^m  \otimes Q\sub{\aa}^\mu \varrho_i Q\sub{\aa}^\mu) U^\dagger], \nonumber \\
\<O\sub{\s}\>_\mathrm{before}^{\rho, M_i, x} &= \frac{1}{p_\rho^M(x)} \sum_{l}\sum_{(m,\mu)_l} \tr[(\one\sub{\s} \otimes P\sub{\aa}^x ) U (O\sub{\s} Q\sub{\s}^m\rho Q\sub{\s}^m  \otimes Q\sub{\aa}^\mu \varrho_i Q\sub{\aa}^\mu) U^\dagger],
\end{align}
which results in \eq{eq:exp-O-symmetric}. 
\end{proof}

\end{document}